\documentclass[aps,pra,superscriptaddress,reprint,amsmath,amssymb,showpacs,floatfix]{revtex4-1}

\usepackage{amsfonts}
\usepackage{amsmath}
\usepackage{fixmath}
\usepackage{graphicx}
\usepackage{cancel}
\usepackage{color}
\usepackage{url}
\usepackage[bookmarks=false]{hyperref}
\usepackage{slashed,wrapfig,nicefrac}
\usepackage{amssymb}
\usepackage{pifont}
\usepackage{floatrow}

\usepackage{comment}
\usepackage[usenames,dvipsnames]{xcolor}

\usepackage{silence}
\renewcommand\vec{\mathbold}

\newcommand\ket[1]{\left|#1\right\rangle}

\begin{document}

\preprint{APS/123-QED}

\title{Phase Diagram of a Strongly Interacting Spin-Imbalanced Fermi Gas}
\author{Ben A. Olsen}
\affiliation{Department of Physics \& Astronomy and Rice Center for Quantum Materials, Rice University, Houston, TX 77005, USA}
\author{Melissa C. Revelle}
\affiliation{Department of Physics \& Astronomy and Rice Center for Quantum Materials, Rice University, Houston, TX 77005, USA}
\author{Jacob A. Fry}
\affiliation{Department of Physics \& Astronomy and Rice Center for Quantum Materials, Rice University, Houston, TX 77005, USA}
\author{Daniel E. Sheehy}
\affiliation{Department of Physics \& Astronomy, Louisiana State University, Baton Rouge, LA 70803, USA}
\author{Randall G. Hulet}
\affiliation{Department of Physics \& Astronomy and Rice Center for Quantum Materials, Rice University, Houston, TX 77005, USA}
\email{randy@rice.edu}

\date{\today}

\begin{abstract}
We obtain the phase diagram of spin-imbalanced interacting Fermi gases from measurements of density profiles of $^6$Li atoms in a harmonic trap. 
These results agree with, and extend, previous experimental measurements. 
Measurements of the critical polarization at which the balanced superfluid core vanishes generally agree with previous experimental results and with quantum Monte Carlo (QMC) calculations in the BCS and unitary regimes.
We disagree with the QMC results in the BEC regime, however, where the measured critical polarizations are greater than theoretically predicted.
We also measure the equation of state in the crossover regime for a gas with equal numbers of the two fermion spin states.
\end{abstract}

                              
\maketitle

Strongly interacting Fermi gases are found in a variety of settings, including superfluid $^3$He, quark matter, superconducting materials, and ultracold  atomic gases~\cite{giorginireview, sheehyreview, chevy}.
The properties of such systems, including the nature of any superfluid or superconducting order, strongly depend on the interactions between particles. 
At sufficiently low temperatures the short-range interaction between opposite spin atomic fermions may be characterized by the parameter $1/k_Fa$, where $k_F$ is the Fermi momentum, and $a$ is the $s$-wave scattering length.
For $1/k_Fa \gtrsim 1$, opposite spins may form tightly bound bosonic pairs which repel each other, thus creating a Bose-Einstein condensate (BEC) of molecules.
For weaker attraction, where $1/k_Fa \lesssim -1$, an ultracold atomic gas may form a conventional Bardeen-Cooper-Schrieffer (BCS) superfluid of loosely bound pairs.
In between these extremes is the unitarity regime, $-1 < 1/k_Fa < 1$, corresponding to resonant two-body interactions.
This BEC--BCS crossover has been studied extensively over the past decade in the context of ultracold atomic Fermi gases~\cite{zwerger,levin12,mohit}.

When the two spin states have equal populations, the crossover between the BEC and BCS limits has no phase transitions as a function of $1/k_Fa$.  
Additional phases can appear, however, when an effective magnetic field couples to the spin-$\nicefrac{1}{2}$ fermions, favoring an imbalance (or polarization) in the number of fermions in each spin state~\cite{sheehy07,gubbels13}. 
In thin-film electronic superconductors, such a coupling can come from a
real in-plane magnetic field~\cite{Wu1994}.  
In the present setting of cold atomic gases, this imbalance is accomplished by creating unequal populations of the two hyperfine levels comprising the pseudo-spin-$\nicefrac{1}{2}$ system.  
In the BCS regime, a sufficiently large chemical potential difference, known as the Chandrasekhar-Clogston (CC) limit~\cite{chandrasekhar62, clogston62}, will suppress pairing.
A spin imbalance can be accommodated in the BEC regime, however, resulting in a Bose-Fermi mixture that remains a superfluid.
The exotic Fulde-Ferrell--Larkin-Ovchinnikov (FFLO) state, featuring pairs with non-zero momentum, has been proposed as the ground state of a spin-imbalanced superconductor under certain conditions~\cite{FF, LO}. There have been no definitive observations of FFLO superconductivity, but an experiment on  spin-imbalanced fermions confined to one dimension has produced a phase diagram with a large polarized region consistent with FFLO~\cite{liao10}.

Neglecting any exotic superfluid phases (such as the FFLO), the phase diagram of  the three dimensional (3D) spin-imbalanced Fermi gas as a function of interaction strength and polarization, exhibits  four phases~\cite{SR2006}:  (i) fully polarized, non-interacting normal (N$_\text{FP}$), (ii) partially polarized normal (N$_\text{PP}$), (iii) partially polarized superfluid (SF$_\text{P}$), and (iv) unpolarized superfluid (SF$_0$)~\cite{pilati08}.
Additionally, constraining the system to fixed particle number leads to regions of phase-separated mixtures of these phases. 
The local polarization, defined as the effective magnetization divided by the density, $p=(n_\uparrow - n_\downarrow)/(n_\uparrow + n_\downarrow)$, vanishes in the SF$_0$ phase, $p=1$ in the N$_\text{FP}$ phase, and $0<p<1$ in the N$_\text{PP}$ and SF$_\text{P}$ phases.
The majority and minority species are defined by $n_\uparrow \geq n_\downarrow$.

Experimentally, atoms are generally trapped in potentials resulting in inhomogeneous density distributions.  
In the local density approximation (LDA) the local state of the gas is determined by its local chemical potential, so the density profiles can reveal transitions between phases.  
Observations of phase separation in spin-imbalanced Fermi gases were obtained at Rice and MIT by direct \emph{in-situ} imaging of the density distributions~\cite{partridge06, partridge06b, shin06}, and by imaging the distributions in time-of-flight~\cite{zwierlein06b}.  
The distributions in the Rice experiment were out-of-equilibrium due to an evaporative depolarization mechanism at work in their highly elongated confining potential~\cite{parish09,liao11}, and could not, therefore, be compared with distributions calculated assuming equilibrium.   
Density profiles obtained by the MIT group at unitarity and on the BEC side of resonance~\cite{shin08} agree quantitatively with the theory of Bertaina and Giorgini (BG) computed using quantum Monte Carlo (QMC) and the LDA~\cite{bertaina09}.  
In the unitary regime, these profiles contain a jump in the local polarization $p$ that indicates a first-order phase transition between the superfluid and the normal phases~\cite{shin08, bertaina09}.  
The ENS group measured thermodynamic properties of the imbalanced gas by extracting the equation of state from doubly-integrated density profiles~\cite{navon10}.  
In this paper, we report new measurements of the density profiles of Fermi gases for $-1 \lesssim 1/k_Fa \lesssim 2$ and use these measurements to better constrain the low-temperature phase diagram.  
These measurements largely confirm the results of previous investigations and extend the range of interactions studied.


\begin{figure*}[ht]
\centering
{\includegraphics[width=\textwidth]{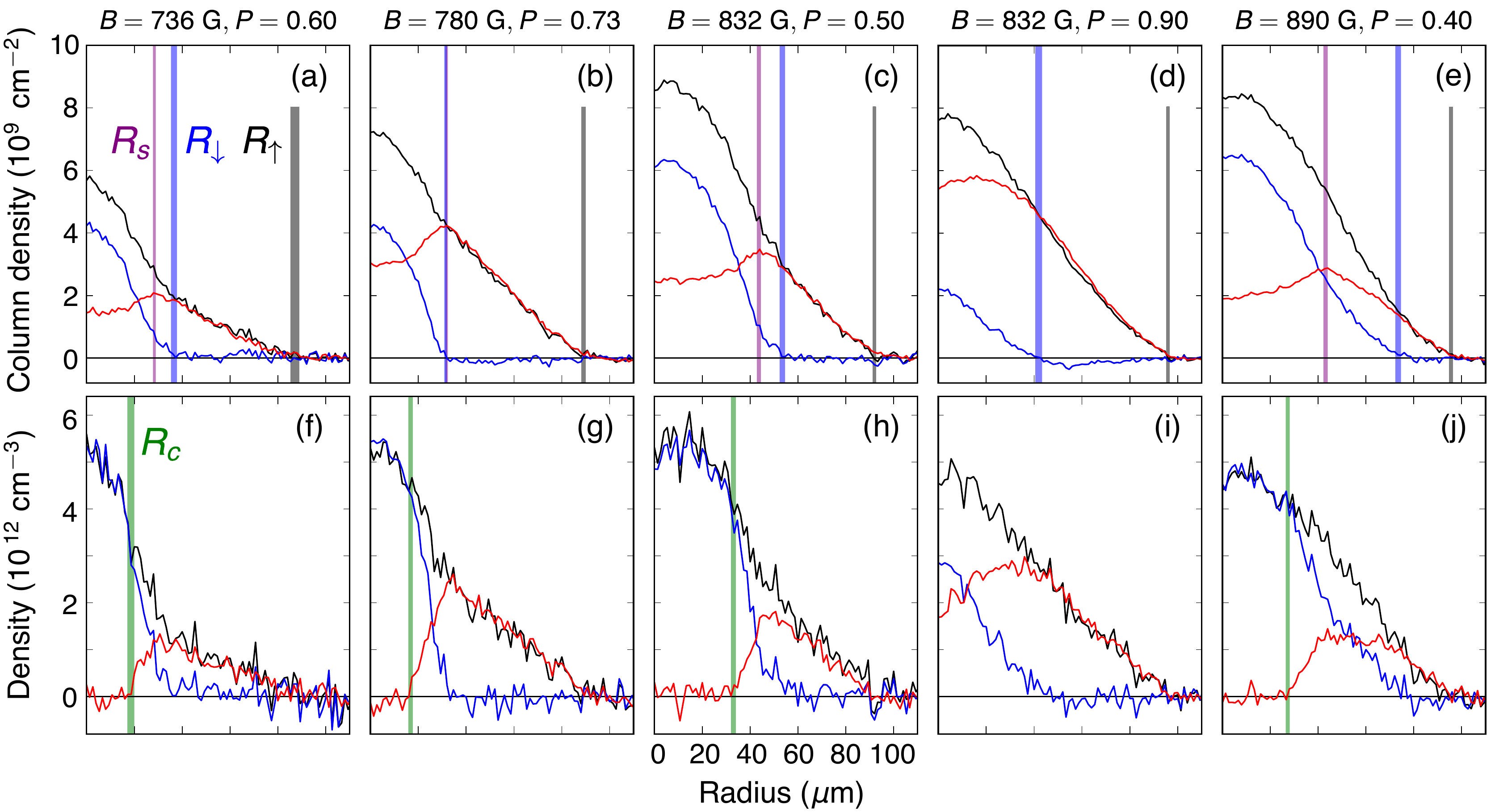}}
{\caption{(Color online) (a)--(e) Axial cuts of the quadrant-averaged column density $n_{cq\uparrow,\downarrow}(z)$ and (f)--(j) their corresponding density profiles $n_{\uparrow,\downarrow}(z)$. 
Values of $B$ and $P$ are indicated for each column, and the corresponding values of $1/k_{F\uparrow}a$ are: (a, f) $1.6$; (b, g) $0.6$; (c, h), (d, i) $0$; (e, j) $-0.4$.
The uncertainty in $1/k_{F\uparrow}a$ is as large as $0.07$ based on a combination of a 10\% systematic uncertainty and shot-to-shot variation in $N_\uparrow$, 3\% uncertainty in the trap frequencies, and 2~G uncertainty in the bias magnetic field.
Each plot is an average of 3--9 experimental realizations that have $P$ within a range $\Delta P =0.02$ centered on the given value. 
Black, blue and red curves correspond to the majority spin $\ket \uparrow$, minority spin $\ket \downarrow$, and their difference, respectively. 
In the upper row, the blue and black vertical lines indicate the mean of the minority and majority edges, $R_\downarrow$ and $R_\uparrow$, respectively, and the purple vertical lines indicate the radius of maximum column density difference $R_s$. 
In the lower row, the vertical green lines indicate the mean boundary of the SF$_0$ core, $R_c$. 
For each vertical line, the standard error of the mean is indicated by the line's thickness.
We estimate a systematic uncertainty in the radii of $4~\mu$m, dominated by the resolution limit of our imaging system.
For (d, i), $P>P_c$ so that $R_c=0$, and $R_s$ is not meaningful.}
	\label{profile}}
\end{figure*}


Our method for producing an imbalanced degenerate gas in the lowest two hyperfine states of $^{6}$Li, $F = 1/2, \, m_F = 1/2$ ($\ket\uparrow$) and  $F = 1/2, \, m_F = - 1/2$ ($\ket\downarrow$), has been discussed previously in detail~\cite{partridge06,partridge06b, liao10}. 
In brief, we sympathetically cool $^6$Li with $^7$Li in an Ioffe-Pritchard magnetic trap, then load the $^6$Li into a single-beam optical dipole trap formed by a focused infrared laser beam. 
We control the spin imbalance by varying the power of an adiabatic RF transfer from $\ket{\uparrow}$ to $\ket{\downarrow}$ at a field of 835~G. 
After the RF transfer, we evaporatively cool the cloud in the single-beam trap by reducing its depth. We evaporate at 835~G to study interactions on the BCS side of the broad Feshbach resonance at 832~G~\cite{houbiers98, zurn13}, while for fields on the BEC side of resonance we quickly ramp the field to 765~G before evaporation.
After evaporation, atoms are loaded into the final trap formed by two focused infrared laser beams crossing at right angles while the single-beam trap is slowly ($100$~ms) ramped off. 
The crossed beams each have $1/e^2$ radii of $55~\mu\text{m}\times235~\mu\text{m}$, resulting in an ellipsoidal crossed-beam trap with a measured axial frequency of $\omega_z/2\pi = 78$~Hz and measured radial frequencies of $\omega_x/2\pi=248$~Hz and $\omega_y/2\pi=274$~Hz, at a trap depth of 1.5~$\mu$K. 
The number of $\ket \uparrow$ atoms, $N_\uparrow$, is typically around $2\times10^5$, and varies by about 10\% shot-to-shot.
The cloud polarization $P = \frac{N_\uparrow - N_\downarrow}{N_\uparrow+N_\downarrow}$ varies from shot-to-shot by about 30\% for a given RF power, so data must be post-selected using the measured $P$. 
After loading into the crossed-beam trap, we ramp the magnetic field to its final value $B$ at a rate from 0.4--2.0~G/ms; the final bias field has an uncertainty of 2~G. 

We use {\it in situ} phase-contrast polarization imaging, described previously~\cite{liao10, bradley97} to record the spatial distribution of the trapped atoms.
The probe beam propagates perpendicular to the bias magnetic field, which is parallel to the axial trap direction. 
The column densities, $n_{c\uparrow,\downarrow}(x',z)$, of each spin state are extracted from two images taken within several $\mu$s of each other at different probe detunings. Here the imaging plane $(x',z)$ is rotated 30$^\circ$ from the $(x,z)$ plane defined by the trap potential.
To improve the signal-to-noise ratio, we fit $n_c$ to find the cloud center $(x'=0,z=0)$, then average the four quadrants to obtain the column density distributions of the majority, $n_{c q\uparrow}(x',z) = \frac{1}{4}[n_{c \uparrow}(x',z)+n_{c \uparrow}(-x',z)+n_{c \uparrow}(x',-z)+n_{c \uparrow}(-x',-z)]$, minority $n_{cq \downarrow}(x',z)$, and their difference, which is related to the spin density. 
The top row of Fig.~\ref{profile} shows the average of these column densities for several experimental realizations with fixed parameters, for several values of $B$ and $P$.
The majority and minority cloud radii, $R_\uparrow$ and  $R_\downarrow$, are obtained from axial cuts of the column densities for each experimental run, then averaged over several runs.
We also determine the radius, $R_s$, where the spin column density, $n_{cq\uparrow}-n_{cq\downarrow}$, is maximum (the `cusp'). 
Within the LDA, a cusp with a discontinuous derivative would indicate the location of a first-order phase transition for a \emph{uniform} gas.
These mean radii are indicated by the vertical lines in Fig.~\ref{profile}.

\begin{figure*}[t]
	\centering
		\includegraphics[width=\textwidth]{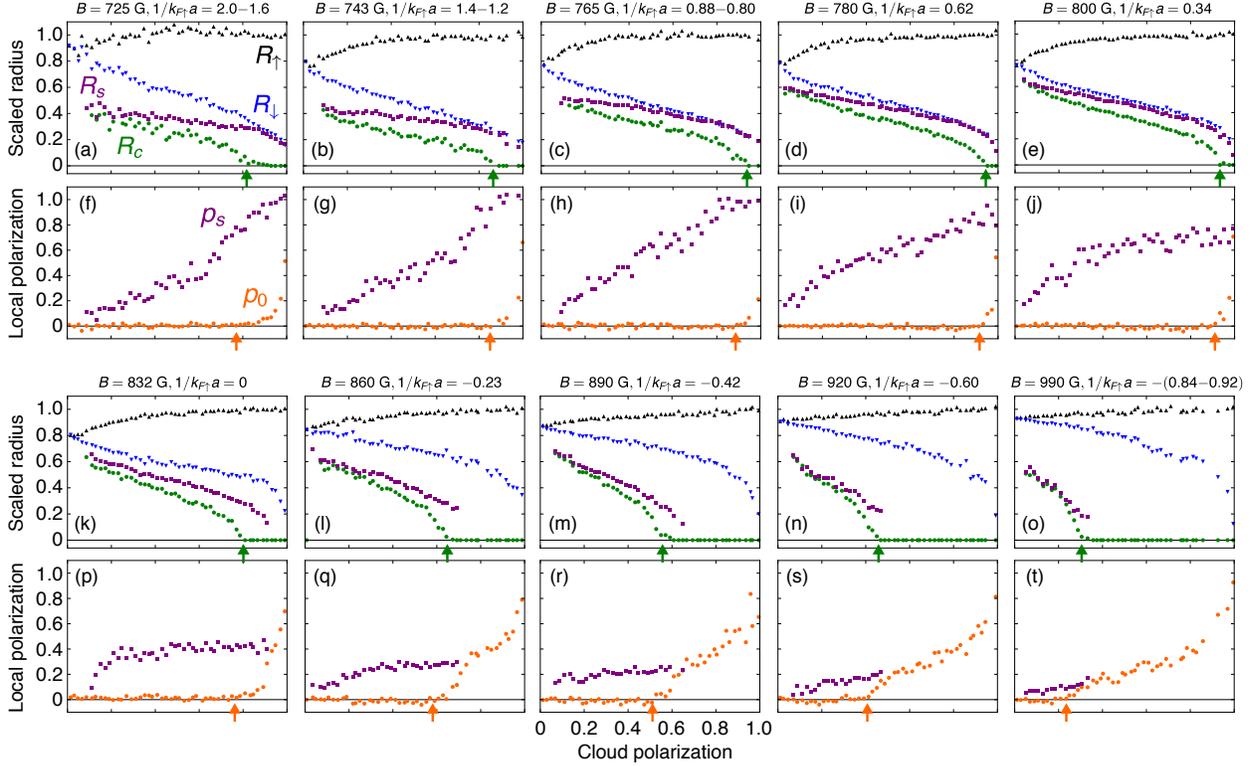}
	\caption{(Color online) (a)--(e), (k)--(o) Radii extracted from density and column density profiles at several interaction strengths: the majority $R_\uparrow$, ($\blacktriangle$), minority $R_\downarrow$ ($\textcolor[rgb]{0,0,1}{\blacktriangledown}$), cusp $R_s$ (\textcolor{Purple}{\scriptsize$ \blacksquare$}), and SF$_0$ core $R_c$ ($\textcolor{Green}{\bullet}$) radii as functions of $P$, scaled by the axial Thomas-Fermi radius, $R_z$, of a non-interacting Fermi gas with $N_\uparrow$ particles. 
(f)--(j), (p)--(t) Local polarization at the cloud center, $p_0$ ($\textcolor{Orange}{\bullet}$), and at $R_s$, $p_s$ (\textcolor{Purple}{\scriptsize$\blacksquare$}). 
Each data point is the average of several realizations of the experiment, binned with width $\Delta P=0.02$. 
Some of the phase boundaries $R_s$ and $R_c$ could not be identified for small $P$ due to poor signal-to-noise, and for high $P$ there is no identifiable $R_s$. 
In these instances, the data points are omitted. 
The values for $1/k_{F\uparrow}a$ have uncertainty less than $0.07$, resulting from 10\% systematic uncertainty and shot-to-shot variation in $N_\uparrow$, 3\% uncertainty in the trap frequencies, and 2~G uncertainty in the bias magnetic field. 
However, due to systematic variation of $N_\uparrow$ with $P$, $1/k_{F\uparrow}a$ varies with $P$ for a given $B$, particularly in the deep BCS and BEC regimes.
In these cases, we list a range of values from $1/k_{F\uparrow}a$ at $P=0$ to the value at $P=1$, otherwise, we list the mean value.
For each interaction strength, $R_c$ decreases as $P$ increases, until vanishing at $P_{c}$ (green arrow), which we determined with a fit (see text). 
We also fit $p_0$ to determine $P_c$ (orange arrows), as described in the text. }
	\label{phases}
\end{figure*}

We reconstruct the density distributions $n_{\uparrow,\downarrow}(\vec r)$  using inverse Abel transforms of the averaged $n_{cq\uparrow,\downarrow}$. 
The bottom row in Fig.~\ref{profile} shows axial cuts of these density distributions. 
The SF$_0$ core radius, $R_c$, is the radius at which the spin density first rises above zero.
The mean radii for several experimental realizations are indicated by the vertical lines in the bottom row of Fig.~\ref{profile}. 
Experimentally, we determine $R_c$ by finding where the spin density first rises above the background spin density noise, which is the standard deviation of the spin density for $z>R_\uparrow$. 
To reduce bias toward obtaining smaller values of $R_c$ due to noise, we smooth the profiles with a 7-pixel-wide Hann window before computing $R_c$.
We also confirm our determination of $R_c$ by fitting the spin density profiles near $R_c$ with a function that increases linearly from $0$ for $z>R_c$; the fit results are consistent to within shot-to-shot variation.

Temperatures are measured by fitting the ferromagnetic wings of $n_{c\uparrow}$ for clouds with high $P$ to non-interacting Thomas-Fermi distributions. 
We find that for $B\geq 743$~G the fitted temperature $T\lesssim0.08\, T_F$, where $T_F\approx 1.5~\mu$K is the Fermi temperature of $N_\uparrow$ non-interacting atoms. 
For lower values of $B$, however, we measure higher temperatures, which are likely a result of heating from inelastic molecular decay collisions. 
At $B=725$~G, for example, we find $T\approx0.11\,T_F$.

The boundary locations $R_\uparrow$, $R_\downarrow$, $R_s$, and $R_c$, are plotted as functions of $P$ in Fig.~\ref{phases}, for several different interaction strengths, $1/k_{F\uparrow}a$, ranging from the BEC to the BCS regimes.  
These boundary radii are normalized by $R_z=(48N_\uparrow)^{1/6}a_z(\omega_x\omega_y/\omega_z^2)^{1/6}$, the axial Thomas-Fermi radius for a non-interacting gas with $N_\uparrow$ atoms, where $a_z=(\hbar/m\omega_z)^{1/2}$ is the axial harmonic oscillator length. 
The interaction strength is determined from $k_{F\uparrow} = (48 N_\uparrow)^{1/6}/\bar a_\text{ho}$ and $a=a(B)$~\cite{zurn13}, where $\bar a_\text{ho}=(\hbar^3/m^3\omega_z\omega_x\omega_y)^{1/6}$ is the mean harmonic oscillator length, and $B$ is the bias magnetic field. 
For a given $B$, the systematic variation in $N_\uparrow$ with $P$ produces up to a factor of $1.2$ variation in $1/k_{F\uparrow}a$.
Due to this variation, experiments at a given field trace out the $P$--$1/k_{F\uparrow}a$ phase diagram along non-vertical lines. To account for day-to-day variation in trap frequencies, we scale $R_z$ for all the data at a given $B$ so that $R_\uparrow/R_z$ goes to 1 as $P$ goes to 1---this variation is less than 5\%.

The radii plotted in Fig.~\ref{phases} provide detailed information about the phases of trapped imbalanced Fermi gases as a function of the imposed population imbalance.
One common feature is the existence of a balanced SF$_0$ core with radius $R_c$ that decreases with increasing $P$ until it vanishes at a critical cloud polarization, $P_{c}$. 
To extract $P_c$ we fit $R_c(P)$ for each field, shown by the green data points in Fig.~\ref{phases}, to an empirical function which vanishes as $(P_c-P)^{1/2}$ for $P<P_c$. 
The results are indicated by vertical green arrows in Fig.~\ref{phases}. 
At unitarity, we measure $P_{c} = 0.79(4)$, where the error bar accounts for the uncertainty in measuring $P$ for a single cloud as well as systematic uncertainty in the best fit parameters. 
This result is in good agreement with previous measurements giving $P_{c} = 0.77$~\cite{shin06}, $0.76(3)$~\cite{nascimbene09}, and $0.75$~\cite{navon10}, and with theoretical predictions of $P_c=0.77$~\cite{bertaina09,lobo06}, all slightly higher than an initial measurement of $P_{c} = 0.70(3)$~\cite{zwierlein06b}. 

The work of BG, following earlier calculations of Pilati and Giorgini~\cite{pilati08}, involved calculating the phase diagram of trapped Fermi gases by combining the LDA with fits to QMC calculations to characterize the ground-state energies of the strongly interacting balanced SF$_0$ phase and the partially polarized normal phase N$_\text{PP}$.
The ground-state energy of the SF$_P$ was taken to consist of contributions from a balanced superfluid of pairs (given by the SF$_0$ equation of state),  a noninteracting Fermi gas of the excess $\ket \uparrow$ spins, and a leading-order interaction between $\ket \uparrow$ spins and pairs characterized by the atom-pair scattering length $a_{bf} = 1.18a$.  
This characterization of the SF$_\text{P}$ state was found by Pilati and Giorgini to agree quite well with their QMC calculations.  
In addition, we have repeated the BG calculations including additional terms in the expression for the ground-state energy.  
The theory of BG includes an interaction between Cooper pairs, with density $\propto n_\downarrow$, and excess $\ket \uparrow$ spins, with density $\propto n_\uparrow - n_\downarrow$, resulting in an interaction strength $\propto n_\downarrow ( n_\uparrow - n_\downarrow)$.
Work by Alzetto and Leyronas has found a higher-order correction with strength $\propto n_\downarrow( n_\uparrow - n_\downarrow)^{4/3}$ ~\cite{Alzetto}.  
However, we find that including this correction (contained in Eq.~(53) of Ref.~\cite{Alzetto}) within the BG formalism does {\em not} appreciably alter the value of $P_c$ for any interaction strength.
Thus, we expect the BG result for $P_c$ as a function of interaction strength to be a robust theoretical prediction that we can test with our measurements.  

We also determine $P_c$ by finding the value of $P$ where the local polarization at the center of the cloud, $p_0=p(z=0)$, first rises above zero by fitting $p_0(P)$ to a function that increases with $P$ for $P>P_c$.
For $1/k_{F\uparrow}a>0.5$, where we find a continuous SF$_0$--SF$_\text{P}$ phase boundary, we assumed that $p_0(P)$ increases with a sum of terms going like $(P-P_c)^{3/2}$ and $(P-P_c)^{5/2}$.  
This form is motivated by the mean-field result for the magnetization $M$ vs.\ chemical potential difference in the SF$_\text{P}$ state of a 3D Fermi gas~\cite{sheehy07},
\begin{equation}
\label{magnetization}
M = \frac{2}{3}\frac{m^{3/2}}{\sqrt{2}\pi^2 \hbar^3}\Big( \sqrt{h^2-|\Delta|^2}-|\mu|\Big)^{3/2}\Theta(h-h_c),
\end{equation}
with $m$ the atom mass, $h$ the chemical potential difference, $\mu$ the chemical potential and $\Delta$ the local pairing amplitude.  
As seen by the presence of the Heaviside step function $\Theta(h-h_c)$, the magnetization is nonzero only for $h>h_c  = \sqrt{|\mu|^2 +\Delta^2}$ and, close to $h_c$ the onset of $M$ is a sum of terms going as $(h-h_c)^{3/2}$ and $(h-h_c)^{5/2}$.  
If we furthermore assume that,
at low $P$, the cloud polarization scales linearly with $h$, then we have justified our assumed form for the behavior of $p_0(P)$, allowing us to extract $P_c$.

Away from the deep BEC regime, for $1/k_{F\uparrow}a<0.5$, where we find a
first-order phase transition
SF$_0$$\rightarrow$N$_\text{PP}$~\cite{bertaina09}, we fit  $p_0(P)$ with a
function that is linear in $(P-P_c)$, the expected magnetization for a Pauli
paramagnetic phase.  The values of $P_c$ obtained from these fits 
 are indicated by vertical
orange arrows in Fig.~\ref{phases}.  While our two methods should
ideally produce the same $P_c$, they differ slightly because we only
consider non-negative radii, which leads to slight overestimates of
$R_c$ near $P_c$ when averaging several profiles.  The magnitude of
this effect is smaller than the uncertainty in determining $P$.
Furthermore, due to noise in the density profiles, we cannot
distinguish an SF$_0$ core from an SF$_\text{P}$ phase with $p<0.03$.

\begin{figure}[t]
	\centering
		\includegraphics[width=\columnwidth]{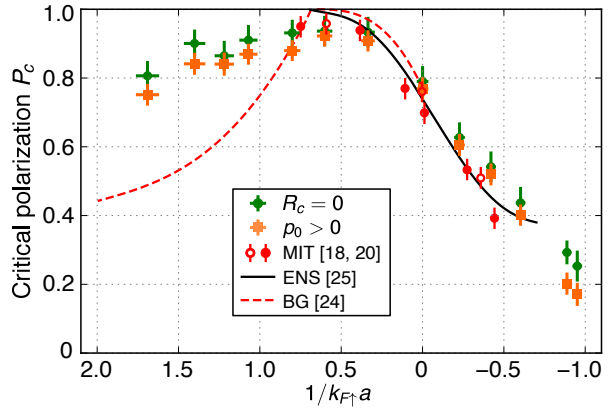}
	\caption{(Color online)  Critical polarization of a trapped gas, $P_{c}$, as a function of the interaction parameter $1/k_{F\uparrow}a$ in the BEC-BCS crossover. 
An unpolarized superfluid core exists for $P<P_c$. 
The green points are the value of $P$ at which the SF$_0$ core radius vanishes based on the fits described in Fig.~\ref{phases}. 
The orange points show the value of $P$ above which the polarization at the center of the cloud is nonzero based on fitting to the appropriate function (see text). 
Vertical error bars include the uncertainty in determining $P$ of $0.03$, measured by preparing a series of known balanced clouds and finding the variation of $P$, as well as uncertainties in fitted parameters. 
From unitarity to the BCS side, our results agree with previous experimental results from MIT~\cite{,shin06,zwierlein06b}, (open and closed red circles, respectively)~ and from ENS (black line)~\cite{navon10}, as well as with the theory of BG (red dashed line)~\cite{bertaina09}. 
For $1/k_{F\uparrow}a>0.7$, we find $P_{c}$ to be higher than predicted by BG. }
	\label{harmonicp}
\end{figure}

The dependence of the critical polarization $P_{c}$  on $1/k_{F\uparrow}a$ determined by both methods is shown in Fig.~\ref{harmonicp}. 
$P_{c}$ reaches a maximum near $1/k_{F\uparrow}a = 0.7$ and decreases as the interactions are tuned in either direction. 
Our measured values of $P_{c}$ agree with the values from the MIT~\cite{shin06, zwierlein06b} and ENS~\cite{nascimbene09, navon10} groups for $1/k_{F\uparrow}a \leq 0.75$, where our measurement ranges overlap. 
Our measurements also agree with the zero-temperature BG theory~\cite{bertaina09} in this regime.

According to theory, $P_c$ begins to drop for $1/k_{F\uparrow}a>0.7$, as the BCS pairs transition to more tightly bound, bosonic molecules~\cite{pilati08}. 
As $1/k_{F\uparrow}a$ increases, the superfluid becomes more bosonic in character than fermionic, and since the bosonic superfluid can accommodate free fermions, the SF$_0$ core begins to vanish. 
For $1/k_{F\uparrow}a<0.7$, the transition from SF$_0$ to N$_\text{PP}$ is predicted to be first order, while for $1/k_{F\uparrow}a>0.7$ the transition from SF$_0$ to SF$_\text{P}$ is continuous~\cite{pilati08}. 
In this Bose-Fermi regime, we observe critical polarizations for loss of the unpolarized core to be somewhat higher than predicted by BG~\cite{bertaina09}. 
It is unlikely that this discrepancy is due to the elevated temperatures we obtain in the BEC regime, since $P_c$ is expected to decrease with increasing $T$~\cite{parish07}. Thus, the effect of finite $T$ is to \emph{diminish} the SF$_0$ phase in favor of the SF$_\text{P}$ phase, while we actually observe a more robust SF$_0$ phase. 
Another possible explanation is that the discrepancy arises from the experimental challenge of observing a small central polarization that increases from zero continuously with increasing $P$, rather than as a first-order jump, as in the BCS regime.

As we have discussed, the critical polarization $P_c$ indicates where the balanced superfluid core of a trapped gas disappears.
The data also reveal information about the uniform density phase diagram, assuming the LDA holds.  
To study these phase boundaries, we measure the local polarization $p_s$ at the radius of maximum column density difference, $p_{s}=p(R_s)$, for each cloud.  
According to the LDA, jumps in the atom density as a function of chemical potential in a uniform imbalanced gas lead to jumps in the density profile in the trapped gas.  
These jumps occur at radii of maximum column density difference, implying that $p_s$ can indicate the critical polarization for a first-order phase transition in the uniform system~\cite{bertaina09}.


\begin{figure}[t]
	\centering
		\includegraphics[width=\columnwidth]{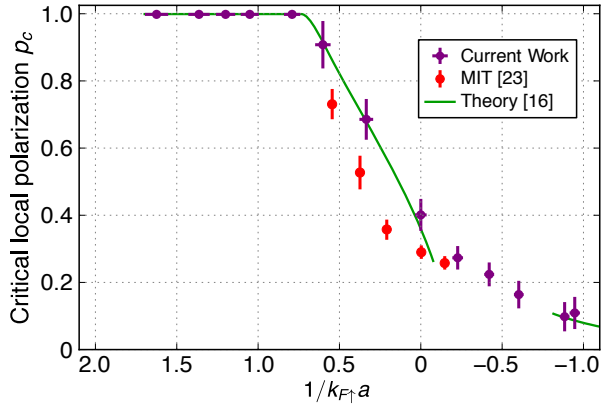}
	\caption{(Color online) Critical local polarization of a homogeneous imbalanced gas, extracted by finding the polarization $p_c$ at the cusp location (where a first order phase transition occurs), as a function of the interaction parameter $1/k_{F\uparrow}a$ in the BEC-BCS crossover. 
Vertical error bars reflect the standard deviation of $p_s(P)$ for $P$ within $\Delta P=\pm0.05$ of $P_c$. 
Our results agree with theory, shown as a green line~\cite{pilati08}, though we find somewhat higher $p_c$ than previous experimental results, indicated by red points~\cite{shin08}.}
	\label{localp}
\end{figure}


In Fig.~\ref{phases}, the second and fourth rows show the dependence of $p_s$ on the cloud polarization $P$.   
We can identify three distinct regimes showing qualitatively different behavior.  
First, in the deep BEC regime, $1/k_{F\uparrow}a> 0.7$, we observe that $p_s$ increases to 1 as $P$ goes to 1.  
This behavior indicates that, in this regime, $p_s$ does not measure the position of a uniform system phase boundary within the LDA, but is instead simply a local maximum of $p$ within an SF$_\text{P}$ phase.  
In this coupling range, therefore, the critical polarization for the superfluid transition of a uniform gas is $p_c=1$~\cite{parish07}.  

Near the unitary region, for $1/k_{F\uparrow}a< 0.3$, $p_s$ is seen to be approximately independent of  $P$ for a wide range of $P$ (see, e.g., Fig.~\ref{phases}p).  
The presence of the plateau indicates that the LDA holds, and that the point of maximum column density difference indeed represents a jump in $p$ and a corresponding phase transition in the uniform system at $p_c$ between SF$_0$ and N$_\text{PP}$ phases. 
We take $p_c$ to be the mean value of the plateau for $P<P_c$.

Finally, in between these two regimes, for $0.3<1/k_{F\uparrow}a<0.7$, we find that $p_s$ increases monotonically with $P$, but that $p_s =1$ is never reached.  
This is the regime, predicted by BG, in which there is an SF$_\text{P}$ phase, but only for sufficiently small $P$.  
Here, we take $p_c$ to be the asymptotic value of $p_s$ evaluated at $P_c$.  
The values of $p_c$ extracted in these regimes are plotted in the phase diagram (Fig.~\ref{localp}), and show excellent agreement with QMC calculations~\cite{pilati08}.


\begin{figure}[hb]
	\centering
		\includegraphics[width=\columnwidth]{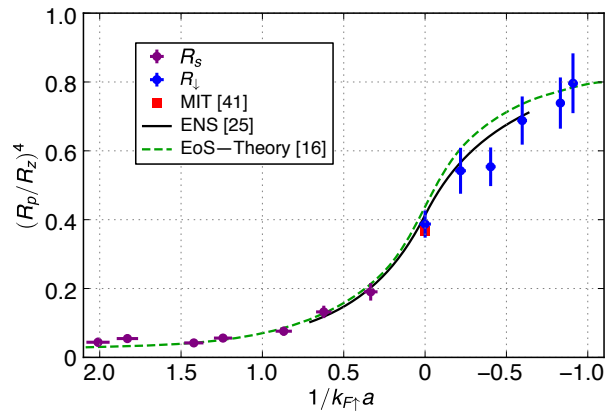}
	\caption{(Color online) The Equation of State (EoS) $\xi(1/k_{F\uparrow}a) = (E_\text{SF}-\frac{N}{2}E_b)/(\frac{3}{5}NE_F)$ for an unpolarized gas. 
The data points show $(R_p/R_z)^4$, where $R_p$ is the superfluid core radius, and $R_z$ is the axial Thomas-Fermi radius of a non-interacting Fermi gas with $N_\uparrow$ atoms. 
At unitarity, and on the BCS side of resonance, we take $R_p=R_\downarrow$ at $P=0$, while on the BEC side, we determine $R_p$ by extrapolating $R_s$ to $P=0$.
Since $N_\uparrow$ varies with $P$ for a given field, the values of $1/k_{F\uparrow}a$ for $P = 0$ differ slightly from those at $P = P_c$, as in Figs.~\ref{harmonicp} and \ref{localp}.  	
 At unitarity, we find $\xi(0)=0.39(3)$ in agreement with a previous measurement from MIT~\cite{ku12} (red point).
Although $(R_p/R_z)^4$  only approximates the EoS away from unitarity, our results agree with theoretical predictions of the EoS~\cite{pilati08} (dashed green line), and with experimental results from ENS~\cite{navon10} (black line).
}
	\label{eos}
\end{figure}


Possible evidence for finite temperature, and perhaps finite imaging resolution, is the absence of clear jumps in the minority density profiles shown in Fig.~\ref{profile} for the unitarity and BCS regimes where a first order transition between SF$_0$ and N$_\text{PP}$ phases is expected.  
Systematic effects are also evident in the phase diagrams of Fig.~\ref{phases}.  
In the unitarity/BCS regimes, $R_s$ should correspond to $R_c$, whereas in the BEC regime for $1/k_{F\uparrow}a\geq 1$, $R_s$ should correspond to $R_\downarrow$, since the transition is between SF$_\text{P}$ and N$_\text{FP}$ phases~\cite{bertaina09}. 
While the predicted trends are observable in the data the agreement is not exact.

Finally, the Equation of State (EoS) of a balanced $(P = 0)$ gas is given by $\xi(1/k_{F\uparrow} a) = (E_\text{SF}-\frac{N}{2}E_b)/(\frac{3}{5} N E_F)$, where $E_\text{SF}$ is the ground state energy of the superfluid, $E_F$ is the Fermi energy, $N$ is the total number of atoms, and $E_b = -\hbar^2/ma^2$ is the binding energy for a molecular pair when $a > 0$~\cite{pilati08,navon10}.  
For a harmonically trapped gas at unitarity, the EoS can be rewritten as $\xi(0) = (R_p/R_z)^4$, where $R_p$ is the radius of the superfluid core and $R_z$ is the Thomas-Fermi radius of a non-interacting gas with $N_\uparrow$ particles~\cite{gehm03,haussmann08}.  
Although $(R_p/R_z)^4$ only approximates the EoS away from unitarity, we nonetheless present our measurements of this quantity in Fig.~\ref{eos}, and compare them with previous measurements and theoretical calculations of the EoS.
In the BEC regime, we fit the column density profiles to a sum of Thomas-Fermi and Gaussian distributions.
Since we find that the Thomas-Fermi radius corresponds to $R_s$ for low $P$, we find $R_p$ by linearly extrapolating $R_s$ to $P = 0$.
At unitarity and in the BCS regime, where the superfluid is unpolarized, we take $R_p = R_\downarrow$ for data with $P = 0$.
At unitarity, we find $\xi(0) = 0.39(3)$, in good agreement with theoretical calculations of the Bertsch parameter~\cite{astrakharchik04,chang04,nishida06,hu06,haussmann07,arnold07}  and recent measurements~\cite{navon10, ku12}.

In conclusion, we have measured density profiles of spin-imbalanced Fermi gases across the BEC-BCS crossover. 
From these profiles, we determined the critical polarization for both harmonically trapped and uniform gases above which the balanced superfluid phase SF$_0$ is suppressed. 
The agreement with previous measurements and QMC theory is generally good, although we find a more robust SF$_0$ core in the BEC regime than predicted by theory.
Although this discrepancy may be explained by very small polarizations that are difficult to detect, the data show that we are able to resolve $p_0$ as small as $0.03$. 
It may also be possible that small adjustments to the theory could result in relatively large changes to $P_c$.
Finally, we have measured the equation of state in the crossover regime, which is consistent with theory from the BCS to the BEC regimes.

\begin{acknowledgments}

The authors would like to thank E. J. Mueller for many stimulating discussions. This work was supported under ARO Grant No. W911NF-13-1-0018 with funds from the DARPA OLE program, NSF, the Welch Foundation (Grant No. C-1133), and ARO-MURI Grant No. W911NF-14-1-0003. In addition, 
D.E.S. was supported by NSF Grant No. DMR-1151717.

\end{acknowledgments}

\end{document}